%% file: bubble.tex
\documentclass{PoS}
\usepackage{lineno}
\usepackage{gensymb}
\usepackage{subcaption}

\title{Fermi Bubbles with HAWC}

\ShortTitle{Fermi Bubbles with HAWC}

\author{\speaker{H. A. Ayala Solares}$^a$, C. M. Hui$^a$ and P. H{\"u}ntemeyer$^a$ for the HAWC Collaboration$^b$\\
        \llap{$^a$}Department of Physics, Michigan Technological University, Houghton, MI, USA\\
        \llap{$^b$}For a complete author list, see \href{http://www.hawc-observatory.org/collaboration/icrc2015.php}{www.hawc-observatory.org/collaboration/icrc2015.php}.\\
        E-mail: \email{hayalaso@mtu.edu}, \email{cmhui@mtu.edu}, \email{petra@mtu.edu} }

\abstract{The Fermi Bubbles, which comprise two large and homogeneous regions of spectrally hard gamma-ray emission extending up to $55^{\circ}$ above and below the Galactic Center, were first noticed in GeV gamma-ray data from the Fermi Telescope in 2010. The mechanism or mechanisms which produce the observed hard spectrum are not understood. Although both hadronic and leptonic models can describe the spectrum of the bubbles, the leptonic model can also explain similar structures observed in microwave data from the WMAP and Planck satellites. Recent publications show that the spectrum of the Fermi Bubbles is well described by a power law with an exponential cutoff in the energy range of 100MeV to 500GeV. Observing the Fermi Bubbles at higher gamma-ray energies will help constrain the origin of the bubbles. A steeper cutoff will favor a leptonic model. The High Altitude Water Cherenkov (HAWC) Observatory, located 4100m above sea level in Mexico, is designed to measure high-energy gamma rays between 100GeV to 100TeV. With a large field of view and good sensitivity to spatially extended sources, HAWC is the best observatory suited to look for extended regions like the Fermi Bubbles at TeV energies. We will present results from a preliminary analysis of the Fermi Bubble visible to HAWC in the Galactic Northern Hemisphere during the ICRC conference.}

\FullConference{The 34th International Cosmic Ray Conference,\\
		30 July- 6 August, 2015\\
		The Hague, The Netherlands}

\begin{document}

\section{Introduction}\input{intro}

\section{Simulation}\input{simulation}


\section{Outlook}\input{conclusion}

\acknowledgments{
\footnotesize{
We acknowledge the support from: the US National Science Foundation (NSF);
the US Department of Energy Office of High-Energy Physics;
the Laboratory Directed Research and Development (LDRD) program of
Los Alamos National Laboratory; Consejo Nacional de Ciencia y Tecnolog\'{\i}a (CONACyT),
Mexico (grants 260378, 55155, 105666, 122331, 132197, 167281, 167733);
Red de F\'{\i}sica de Altas Energ\'{\i}as, Mexico;
DGAPA-UNAM (grants IG100414-3, IN108713,  IN121309, IN115409, IN111315);
VIEP-BUAP (grant 161-EXC-2011);
the University of Wisconsin Alumni Research Foundation;
the Institute of Geophysics, Planetary Physics, and Signatures at Los Alamos National Laboratory;
the Luc Binette Foundation UNAM Postdoctoral Fellowship program.
}}

\bibliographystyle{JHEP}
\bibliography{bubble}

\end{document}

%% file: intro.tex
In 2010, the analysis of \emph{Fermi} data, made by \cite{Dobler2010} and \cite{Su2010}, revealed two large and homogeneous 
regions of spectrally hard gamma-ray emission  extending up to $55^{\circ}$ above and below the Galactic Center. These two regions are called  \emph{Fermi} Bubbles. Since their discovery, multiple models to explain the origin of the bubbles as well as the production of gamma rays have been proposed.
The mechanism or mechanisms which produce the observed hard spectrum of gamma rays are still not understood. Both hadronic and leptonic models can describe the spectrum of the bubbles, however, the leptonic model can also explain similar structures observed in microwave data from the WMAP\cite{Dobler2010}  and Planck satellites\cite{Ade2013}.

The spectrum of the Fermi Bubbles is well described by a power law with an exponential cutoff in the energy range of 100MeV to 500GeV according to \cite{Ackermann2014} and \cite{Yang2014} . Observing the Fermi Bubbles at higher gamma-ray energies will help constrain the origin of the bubbles. A hard spectrum generally points to the presence of hadrons and a cut off spectrum suggests a leptonic origin.

The High Altitude Water Cherenkov (HAWC) Observatory is located 4100m above sea level on the volcano Sierra Negra in the state of Puebla, Mexico and has recently been completed. HAWC is designed to measure high-energy gamma rays between 100GeV to 100TeV. It consists of an array of 300 steel water tanks, called water Cherenkov Detectors (WCDs), covering an area of $\sim22000 \,{\rm m^2}$. Each of the WCDs has four photo-multiplier tubes (PMTs) on the bottom. The PMTs  detect the Cherenkov light in the water that is produced by charged particles from air showers that enter the  WCD. (See \cite{Pretz2015} and \cite{Smith2015} for more details). With a field of view of 2 sr and good sensitivity to spatially extended sources, HAWC is the best observatory to look for extended regions like the Fermi Bubbles at TeV energies. 

In these proceedings we present simulations of the  response to the HAWC detector to two spectral shape assumptions. During the ICRC conference we will present a preliminary analysis on HAWC data to obtain an upper-limit on the flux of the Northern Fermi Bubble.

%% file: simulation.tex
\begin{figure}[!ht]
     \centering
     \begin{subfigure}[b]{0.65\textwidth}
         \includegraphics[width=1\linewidth]{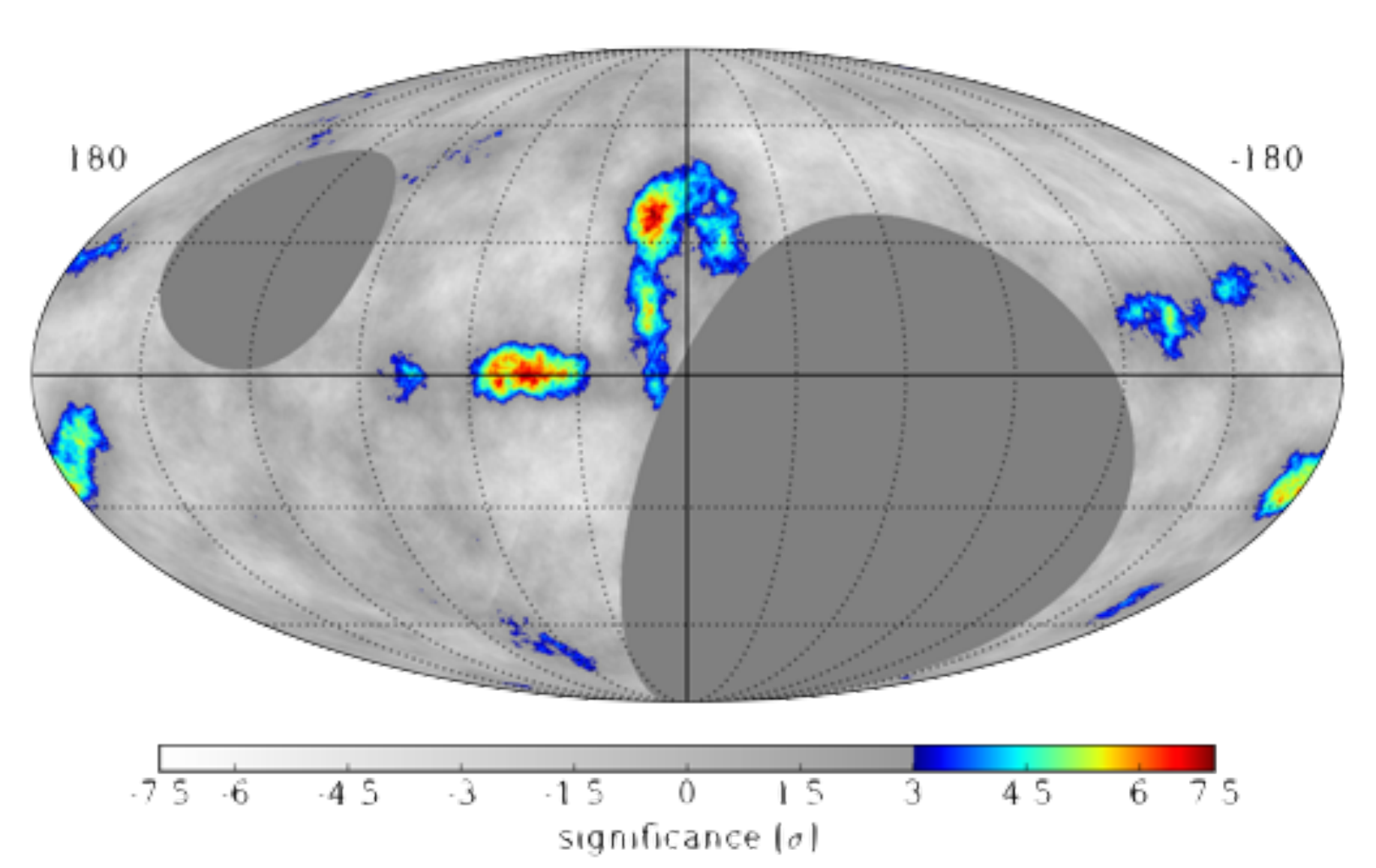}
         \caption{}
         \label{fig:fig1a}
     \end{subfigure}
     \begin{subfigure}[b]{0.65\textwidth}
          \includegraphics[width=1\linewidth]{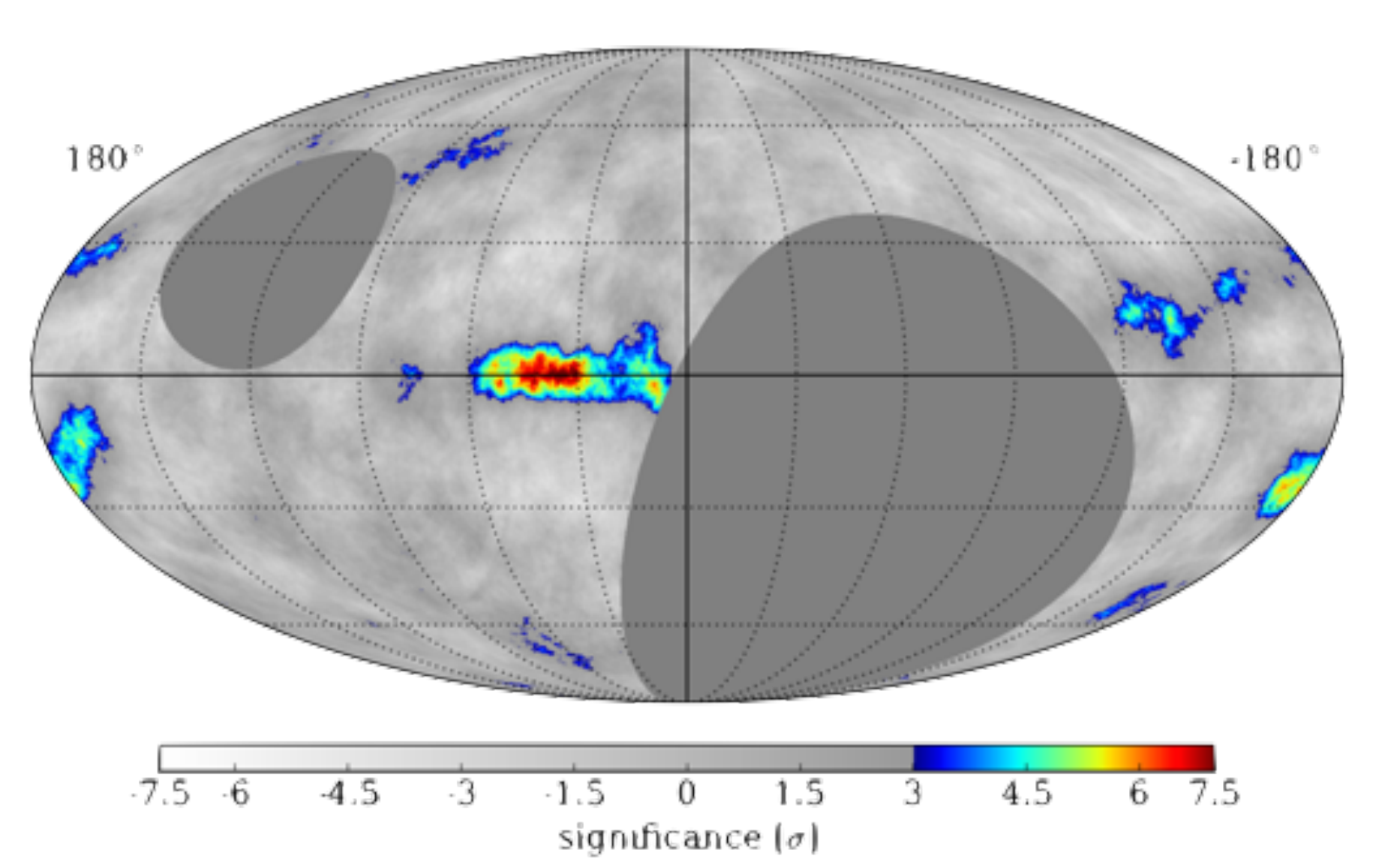}\
          \caption{}
          \label{fig:fig1b}
     \end{subfigure}
     \caption{Significance map for 3 years of simulation shown in Galactic Coordinates; the Galactic plane as well as region A and B (Cosmic rays) are observed. (a)The Fermi Bubbles are simulated with a power-law spectrum. In this case HAWC is able to observe the northern Fermi Bubble. (b) The Fermi Bubbles are simulated with a power-law with cutoff spectrum. HAWC is not able to observe the northern Fermi Bubble in this case.}
\end{figure}

In order to have a first estimate of the response of the HAWC detector to the Fermi Bubbles, a simulation of 3 years of data was performed. The simulation includes cosmic rays, gamma rays from diffuse emission and gamma rays from the Fermi Bubbles. The cosmic rays  follow the spectra of HEAO-3-C2\cite{Engelmann1990}, JACEE\cite{Asakimori1998}, RUNJOB\cite{Apanasenko2001}, TRACER\cite{Ave2008}, ATIC-2\cite{Panov2009}, CREAM-2\cite{Ahn2010}, and PAMELA\cite{Adriani2011}. The gamma rays from diffuse emission come from \textit{GALPROP}\cite{Vladimirov2011}. The HAWC detector configuration for this simulation uses the 300 WCDs.
For the Fermi Bubbles, a simple template with a flat morphology and two spectral assumptions was made. The first assumption is a power-law spectrum with spectral index of 2. The second assumption is a power law with cutoff spectrum. Table \ref{tab:tab1} shows the parameters used for the spectral assumptions (An estimation of these parameters was obtained from \cite{Franckowiak2013}).
\begin{table}[!ht]
\centering
\begin{tabular}{|c|c|c|}
\hline 
Parameters & Power Law & Power Law with cutoff \\ 
\hline 
$E^{2}F_{0} [10^{-7} \rm{GeV} \rm{cm}^{-2} sr^{-1} s^{-1}]$ & 6.31 & 5.10 \\ 
\hline 
Spectral Index & 2.0 & 1.99 \\ 
\hline 
Cutoff Energy (GeV) & --- & 152.48 \\ 
\hline 
\end{tabular} 
\caption{Parameters used for the simulation of the Fermi Bubbles. The differential flux is at 1 GeV}
\label{tab:tab1}
\end{table}
The results of the simulation are shown in figures \ref{fig:fig1a} and \ref{fig:fig1b}.

As it can be seen, with the power law spectrum assumption HAWC would be sensitive enough to observe the Fermi Bubbles. If the spectrum follows a power law with cutoff, HAWC would not be sensitive enough to make a significant detection. 
With either a detection or non-detection, HAWC will be able to  meaningfully constrain the spectrum of the Fermi Bubbles.

%% file: conclusion.tex
We will present an upper limit calculation for the Fermi Bubble using HAWC data. The upper limit calculation will be performed using two spectral assumptions and looking at event excesses in the bubble region. Obtaining the upper limit for specific spectral assumptions will constrain the spectrum of the Fermi Bubbles at $\sim$TeV energies.

%% file: bubble.bbl
\providecommand{\href}[2]{#2}\begingroup\raggedright\begin{thebibliography}{10}

\bibitem{Dobler2010}
G.~Dobler, D.~P. Finkbeiner, I.~Cholis, T.~R. Slatyer, and N.~Weiner, {\it {The
  Fermi Haze: A Gamma-Ray Counterpart to the Microwave Haze}},  {\em ApJ} {\bf
  717} (July, 2010) 825--842, [\href{http://arxiv.org/abs/0910.4583}{{\tt
  arXiv:0910.4583}}].

\bibitem{Su2010}
M.~Su, T.~R. Slatyer, and D.~P. Finkbeiner, {\it {Giant Gamma-Ray Bubbles From
  Fermi -Lat: Active Galactic Nucleus Activity or Bipolar Galactic Wind?}},
  {\em ApJ} {\bf 724} (Dec., 2010) 1044--1082,
  [\href{http://arxiv.org/abs/1005.5480}{{\tt arXiv:1005.5480}}].

\bibitem{Ade2013}
P.~A.~R. Ade et~al., {\it Planck intermediate results.},  {\em A\&A} {\bf 554}
  (2013) A139, [\href{http://arxiv.org/abs/1208.5483}{{\tt arXiv:1208.5483}}].

\bibitem{Ackermann2014}
M.~Ackermann, A.~Albert, et~al., {\it {The Spectrum and Morphology of the Fermi
  Bubbles}},  {\em ApJ} {\bf 793} (2014), no.~1 64.

\bibitem{Yang2014}
R.-z. Yang, F.~Aharonian, and R.~Crocker, {\it {The Fermi Bubbles Revisited}},
  {\em A\&A} {\bf 567} (April, 2014) 8,
  [\href{http://arxiv.org/abs/1402.0403}{{\tt arXiv:1402.0403}}].

\bibitem{Pretz2015}
{\bf HAWC} Collaboration, J.~Pretz, {\it {Highlights from the High Altitude
  Water Cherenkov Observatory}},  in {\em Proc. 34th ICRC}, (The Hague, The
  Netherlands), August, 2015.

\bibitem{Smith2015}
{\bf HAWC} Collaboration, A.~Smith, {\it {HAWC: Design, Operation,
  Reconstruction and Analysis}},  in {\em Proc. 34th ICRC}, (The Hague, The
  Netherlands), August, 2015.

\bibitem{Engelmann1990}
J.~Engelmann, P.~Ferrando, et~al., {\it Charge composition and energy spectra
  of comsic-ray nuclei for elements from be to ni. results from heao-3-c2},
  {\em A\&A.} {\bf 233} (July, 1990) 96--111.

\bibitem{Asakimori1998}
K.~Asakimori, T.~Burnett, M.~Cherry, et~al., {\it Cosmic-ray proton and helium
  spectra: Results from the jacee experiment},  {\em ApJ} {\bf 502} (July,
  1998) 278.

\bibitem{Apanasenko2001}
A.~Apanasenko et~al., {\it Composition and energy spectra of cosmic-ray
  primaries in the energy range $10^{13}$-$10^{15}$ ev/particle observed by
  japanese-russian joint balloon experiment},  {\em APh} {\bf 16} (January,
  2001) 13.

\bibitem{Ave2008}
M.~Ave, P.~Boyle, F.~Gahbauer, et~al., {\it Composition of primary cosmic-ray
  nuclei at high energies},  {\em ApJ} {\bf 678} (May, 2008) 262--273,
  [\href{http://arxiv.org/abs/0801.0582}{{\tt arXiv:0801.0582}}].

\bibitem{Panov2009}
A.~Panov, J.~Adams, J.~H., et~al., {\it Energy spectra of abundant nuclei of
  primary cosmic rays from the data of atic-2 experiment: Final results},  {\em
  Bull. Russ. Acad. Sci. Phys.} {\bf 73} (June, 2009) 564--567,
  [\href{http://arxiv.org/abs/1101.3246}{{\tt arXiv:1101.3246}}].

\bibitem{Ahn2010}
H.~Ahn, P.~Allison, M.~Bagliesi, et~al., {\it Discrepant hardening observed in
  cosmic-ray elemental spectra},  {\em ApJ} {\bf 714} (May, 2010) L89,
  [\href{http://arxiv.org/abs/1004.1123}{{\tt arXiv:1004.1123}}].

\bibitem{Adriani2011}
O.~Adriani et~al., {\it Pamela measurments of cosmic-ray proton and helium
  spectra},  {\em Science} {\bf 332} (April, 2011) 69,
  [\href{http://arxiv.org/abs/1103.4055}{{\tt arXiv:1103.4055}}].

\bibitem{Vladimirov2011}
A.~E. Vladimirov, S.~W. Digel, G.~Johannesson, et~al., {\it Galprop webrun: An
  internet-based service for calculating galactic cosmic ray propagation and
  associated photon emissions},  {\em Elsevier Computer Physics Communications}
  {\bf 182} (2011) 1156--1161.

\bibitem{Franckowiak2013}
A.~Franckowiak and M.~Dmitry, {\it Spectrum and morphology of the fermi
  bubbles},  {\em ICRC} (2013).

\end{thebibliography}\endgroup
